**PAPER • OPEN ACCESS**

# Charmonium production in pp, p+Pb and Pb+Pb collisions with CMS experiment

View the article online for updates and enhancements.

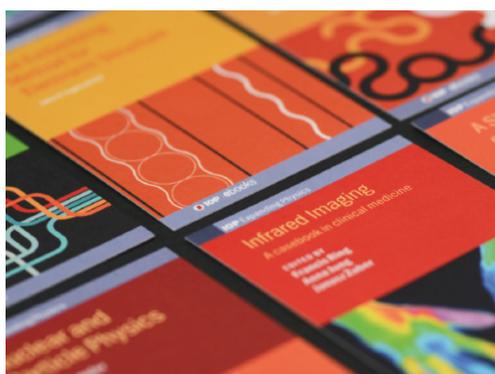

# Charmonium production in pp, p+Pb and Pb+Pb collisions with CMS experiment

**Abdulla Abdulsalam on behalf of the CMS collaboration**

Department of Physics, King Abdulaziz University, Jeddah. KSA

akizhakkepura@kau.edu.sa

**Abstract**. We report the measurements of the prompt and non-prompt J/$\psi$ and $\psi$(2S) nuclear modification factors ($R_{AA}$) using pp and PbPb data and the $R_{pPb}$ using pPb data collected by CMS at the LHC. All measurements were performed via the dimuon decay channel. The $R_{AA}$ analysis is based on PbPb and pp data samples collected at $\sqrt{s_{NN}}$ = 5.02 TeV in 2015, corresponding to integrated luminosities of 464 $\mu$b$^{-1}$ and 28 pb$^{-1}$. These measurements are performed in the dimuon rapidity range of |y| < 2.4 as a function of centrality, rapidity, and transverse momentum ($p_T$) from $p_T$ = 3GeV/c in the most forward region and up to 50 GeV/c. The $R_{pPb}$ analysis based on pPb data, is performed for prompt J/$\psi$ and $\psi$(2S) in different kinematic bins of $p_T$ and rapidity.

## 1. Introduction

Charmonium states constitute an important probe of the Quark-Gluon Plasma (QGP), since they are produced in the early stage of heavy ion collisions and their yields are sensitive to hot QGP medium [1-2]. It has been predicted that the charmonium production can be suppressed in QGP due to the Debye screening of the strong interaction between the charm and anti-charm quarks [3]. The ground states, J/$\psi$ and $\Upsilon$(1S) are expected to be dissociated/melted at higher temperatures than their more-loosely bound excited states. Since the dissociation temperature of the quarkonia depends on their binding energy, they melt at different temperatures in hot QGP and thus the pattern of the quarkonium dissociation can serve as thermometer [4-6] of the medium temperature. Other non-QGP effects (called Cold Nuclear Matter effects) can affect the production of quarkonium in heavy ion collisions and the regeneration process-an additional source of charmonium could enhance the production of the bound states at higher LHC energy [7-8].

　　At the LHC, the inclusive J/$\psi$ meson yield also contains a significant nonprompt contribution coming from b hadron decays. The nonprompt J/$\psi$ component should reflect medium effects on b hadron production in heavy ion collisions, such as b quark energy loss. At the LHC, the ALICE [9] and CMS [10] Collaborations have reported a stronger suppression of the $\psi$(2S) state compared to the J/$\psi$ state in PbPb collisions. In pPb collisions, ATLAS [11] and ALICE [12] data show that $\psi$(2S) suppression, integrated over transverse momentum, is more pronounced than that of the J/$\psi$. In this talk, the measurement of the prompt and non-prompt J/$\psi$ and $\psi$(2S) $R_{AA}$ using PbPb data, collected at the end of 2015 with the CMS experiment at $\sqrt{s_{NN}}$ = 5.02 TeV were presented. Also, measurements of the pPb nuclear modification factor of prompt $\psi$(2S), $R_{pPb}$ in pPb collisions, over the p$_T$ range 4–30 GeV/c and center-of-mass rapidity range –2.4 < y$_{CM}$ < 1.93. The data were collected with the CMS detector at the LHC, in 2013 for the pPb sample and in 2015 for the pp sample at $\sqrt{s_{NN}}$ = 5.02 TeV, corresponding to integrated luminosities of 34.6 ± 1.2 nb$^{-1}$ and 28.0 ± 0.6 pb$^{-1}$, respectively.

## 2. The CMS detector

The central feature of the CMS apparatus is a superconducting solenoid of 6 m internal diameter, providing a magnetic field of 3.8 T. Within the solenoid volume are a silicon pixel and strip tracker, a lead tungstate crystal electromagnetic calorimeter, and a brass and scintillator hadron calorimeter, each composed of a barrel and two endcap sections. Forward calorimeters extend the pseudo-rapidity ($\eta$) coverage provided by the barrel and endcap detectors. The forward hadron (HF) calorimeter uses steel as absorber and quartz fibers as the sensitive material. Together they provide coverage in the range 3.0 < |$\eta$| < 5.2 and serve as luminosity monitors. Muons are detected in gas-ionization chambers embedded in the steel flux-return yoke outside the solenoid, in the range |$\eta$| < 2.4, with detection planes made using three technologies: drift tubes, cathode strip chambers, and resistive-plate chambers. A more detailed description of the CMS detector can be found in Ref. [13].

## 3. Analysis procedure

### 3.1. Signal Extraction

Because of the long lifetime of b hadrons compared to that of J/$\psi$ mesons, the separation of the prompt and nonprompt J/$\psi$ components relies on the measurement of a secondary $\mu^+\mu^-$ vertex displaced from the primary collision vertex. The J/$\psi$ mesons originating from the decay of b hadrons can be resolved using the pseudo-proper decay length $\ell_{J/\psi} = L_{xyz} \cdot m_{J/\psi} \cdot c/|p_{\mu\mu}|$, where $L_{xyz}$ is the distance between the primary and dimuon vertices. To measure the fraction of J/$\psi$ mesons coming from b hadron decays (the nonprompt fraction), the invariant mass spectrum of $\mu^+\mu^-$ pairs and their $\ell_{J/\psi}$ distribution are fitted using a two-dimensional (2D) extended unbinned maximum-likelihood fit. The invariant mass component in the fits was parameterized with the sum of two Crystal Ball functions for the signal, and with a polynomial function for the underlying background. The $\ell_{J/\psi}$ component was parameterized in collision data and Monte Carlo simulated events, using templates for the per-event $\ell_{J/\psi}$ uncertainty distributions, a sum of Gaussian functions to describe the $\ell_{J/\psi}$ resolution, and an empirical combination of exponential functions for the background. Since $\psi$(2S) suffers from lower statistics, 2D fits could not performed with pPb data and prompt charmonia were selected by requiring $\ell_{J/\psi}$ to be smaller than a threshold value [10], which was tuned using simulation in order to keep 90% of the total prompt charmonia. More details of the signal extractions can be found in Ref [10, 14-15].

### 3.2. Acceptance and efficiency corrections

Correction factors were applied to all results to account for detector acceptance, trigger, reconstruction, and selection efficiencies of the $\mu^+\mu^-$ pairs. The corrections were derived from prompt and nonprompt J/$\psi$ meson MC samples in pp, pPb and PbPb, and are evaluated in the same bins of $p_T$, centrality, and rapidity used in the $R_{AA}$ and $R_{pPb}$ analyses. The MC prompt and nonprompt J/$\psi$ meson $p_T$ distributions in rapidity bins were compared to those in data, and the ratios of data over MC were used to weight the MC J/$\psi$ distributions to describe the data better. This weighting accounts for possible mis-modelling of J/$\psi$ kinematics in MC. The systematic uncertainties in the measurements arise from the invariant mass signal and background fitting model assumptions, the parameterisation of the $\ell_{J/\psi}$ distribution, the acceptance and efficiency computation, and sample normalization (integrated luminosity in pp data, counting of the equivalent number of minimum bias events in pPb and PbPb, and nuclear overlap function). These systematic uncertainties are derived separately for pp, pPb and PbPb results, and the total systematic uncertainty is computed as the quadratic sum of the partial terms (more details in Ref [14-15]).

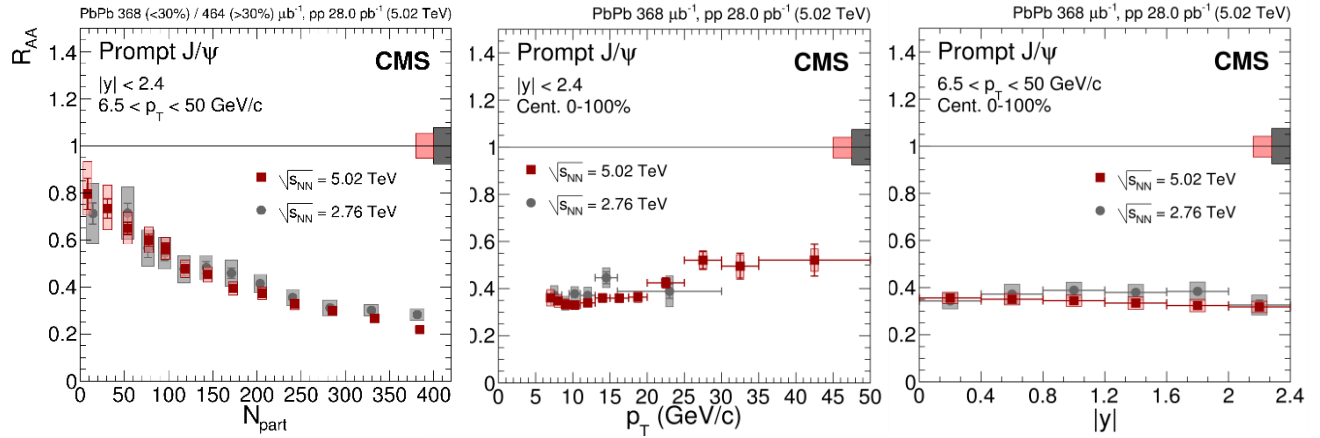

**Figure 1.** $R_{AA}$ of prompt J/ψ mesons as a function of $N_{part}$, $p_T$ and rapidity. The results are compared to those obtained at $\sqrt{s_{NN}}$ = 2.76 TeV

## 4. Results

### 4.1. Charmonium results in PbPb

The figure 1 shows the $R_{AA}$ of prompt J/ψ mesons as a function of $N_{part}$, $p_T$ and rapidity. The results are compared to those obtained at $\sqrt{s_{NN}}$ = 2.76 TeV [16] and they are found to be in good agreement. No strong rapidity dependence of the suppression is observed. As a function of centrality, the $R_{AA}$ is suppressed even for the most peripheral bin (70–100%), with the suppression slowly increasing with $N_{part}$. As a function of $p_T$, the $R_{AA}$ is nearly constant in the range of 5–20 GeV/c, but a lower suppression is observed at higher $p_T$, for the first time in quarkonia. This trend usually attributed to parton energy loss for the charged hadrons in PbPb collisions are seen in $R_{AA}$ at high $p_T$ at $\sqrt{s_{NN}}$ = 5.02 TeV [10]. In the 1.8 < |y| < 2.4 range, the prompt $R_{AA}$ as a function of centrality shows that the suppression is stronger for higher $p_T$ in the most central range. A similar observation by the ALICE Collaboration is attributed to a regeneration contribution [17]. Like the prompt J/ψ, the $R_{AA}$ of nonprompt J/ψ is showing a steady increase of the suppression with increasing centrality of the collision. The $p_T$ dependence of the nonprompt J/ψ shows hints for a smaller suppression at low $p_T$ and hints of a stronger suppression for 1.8 < |y| < 2.4 and $p_T$ > 6.5 GeV/c at all centralities [14-15].

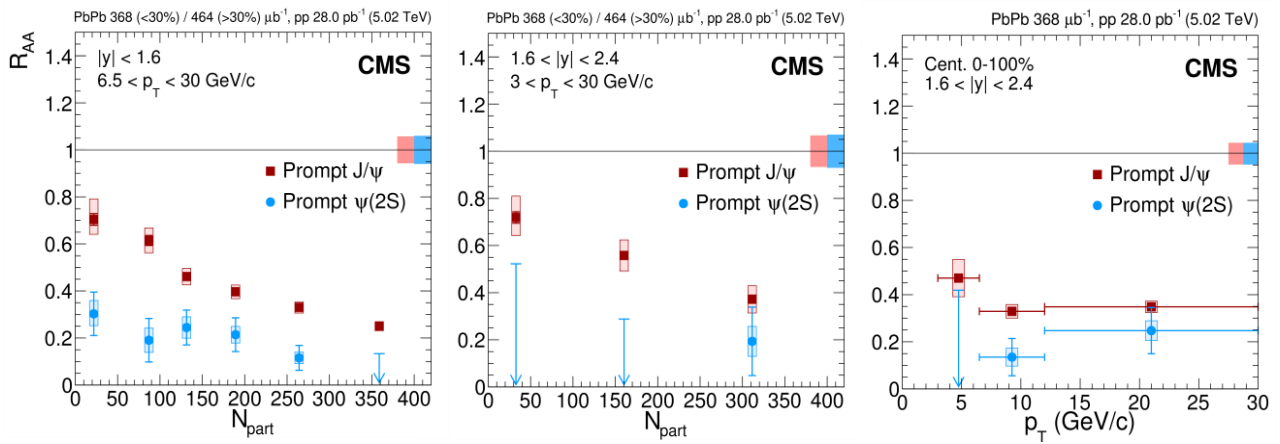

**Figure 2.** $R_{AA}$ of prompt ψ(2S) mesons as a function of $N_{part}$ and $p_T$.

$R_{AA}$ of prompt $\psi(2S)$ were derived by multiplying $R_{AA}$ of prompt J/$\psi$ by the double ratio [($N_{\psi(2S)}$ / $N_{J/\psi}$)$_{PbPb}$ / ($N_{\psi(2S)}$ / $N_{J/\psi}$)$_{pp}$] of the relative modification of the prompt $\psi(2S)$ and J/$\psi$ meson yields from pp to PbPb collisions published in Ref. [10]. The results are presented in the figure 2 as a function of dimuon $p_T$ and centrality. In the bins where the double ratio is not significant, 95% CL intervals on the prompt $\psi(2S)$ $R_{AA}$ are given. The $\psi(2S)$ $R_{AA}$ shows no clear dependence with $p_T$, and hints of an increasing suppression with collision centrality. In the entire measured range, the $\psi(2S)$ production is more suppressed than that of J/$\psi$ mesons, showing that the excited states are more strongly affected by the medium created in PbPb collisions.

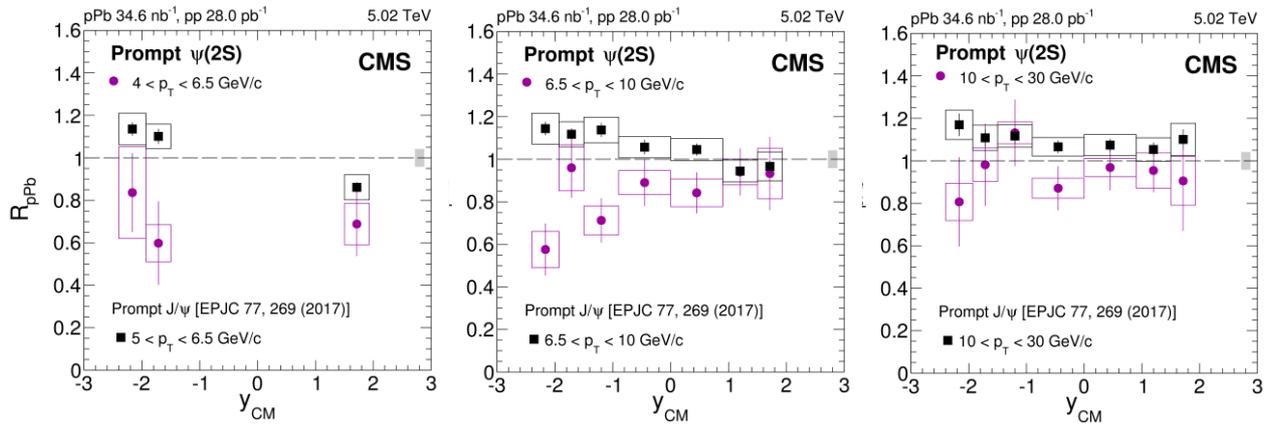

**Figure 3.** Rapidity dependence of the prompt $\psi(2S)$ $R_{pPb}$ in three $p_T$ ranges.

### 4.2. Charmonium results in pPb

The figure 3 shows the rapidity dependence of the prompt $\psi(2S)$ $R_{pPb}$ in three $p_T$ ranges: 4–6.5, 6.5–10, and 10–30 GeV/c. In the lowest two $p_T$ bins, $R_{pPb}$ remains below unity in all rapidity, while in the highest $p_T$ bin, $R_{pPb}$ is consistent with unity. Also, the $R_{pPb}$ of prompt J/$\psi$ [18] is also displayed in the figure for comparison. The J/$\psi$ meson $R_{pPb}$ lies systematically above that of the $\psi(2S)$ state, indicating nuclear effects playing differently in the production of the two states. The figure 4 shows the $p_T$ dependence of the prompt $\psi(2S)$ $R_{pPb}$ in four rapidity bins. The $R_{pPb}$ values in the lowest $p_T$ bins are found to be below unity in all rapidity bins. Over all, there are hints of more suppression of $\psi(2S)$ mesons in all rapidity and $p_T$ bins, unexpected phenomenon in pPb collisions where most of the CNM effects are expected to affect the prompt J/$\psi$ and $\psi(2S)$ mesons in similar ways. However, the fact that the final-state interaction with hadrons moving along with the produced charmonium may lead to a stronger suppression of the $\psi(2S)$ meson, due to its larger size [19], should be considered.

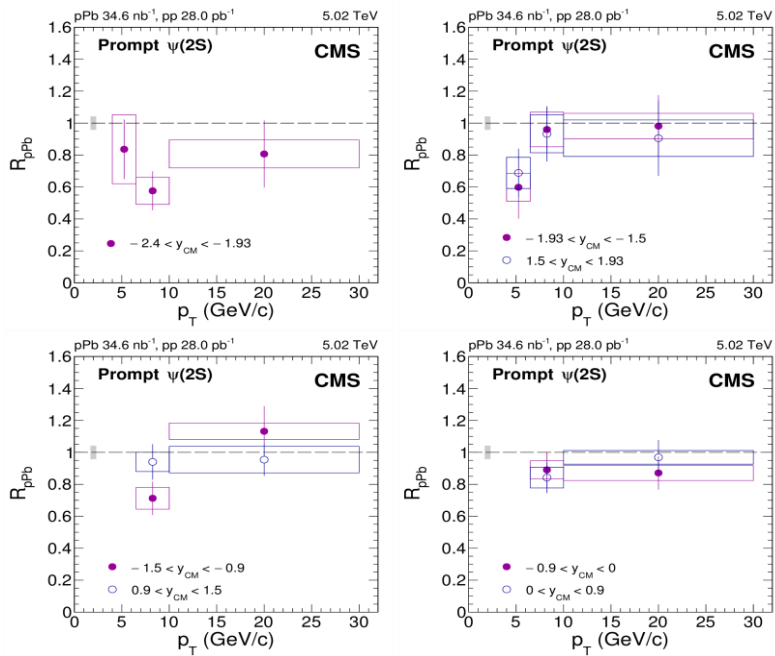

**Figure 4.** Transverse momentum dependence of the prompt $\psi(2S)$ $R_{pPb}$ in four rapidity ranges.

## 5. Summary

The results of the prompt and nonprompt J/$\psi$ meson $R_{AA}$ have been presented in PbPb collisions at $\sqrt{s_{NN}}$ =5.02 TeV, as a function of transverse momentum ($p_T$), and collision centrality in different kinematic ranges. The $R_{AA}$ results show a strong centrality dependence, with an increasing suppression for increasing centrality. For both prompt and nonprompt J/$\psi$ mesons no significant dependence on rapidity is observed. An indication of less suppression in the lowest $p_T$ range at forward rapidity is seen for both J/$\psi$ components. Double-differential measurements show the same trend and suggest a stronger $p_T$ dependence in peripheral events. An indication of less suppression of the prompt J/$\psi$ meson at high $p_T$ is seen with respect to that observed at intermediate $p_T$. The measurements are consistent with previous results at $\sqrt{s_{NN}}$ = 2.76 TeV. The prompt $\psi(2S)$ meson $R_{AA}$ has also been measured in PbPb collisions at 5.02 TeV, as a function of $p_T$ and collision centrality. The results show that the $\psi(2S)$ is more suppressed than the J/$\psi$ meson for all the kinematical ranges studied. No $p_T$ dependence is observed within the current uncertainties. Hints of an increasing suppression with collision centrality are also observed.

In pPb collisions, In the ranges $4 < p_T < 6.5$ and $6.5 < p_T < 10$ GeV/c the value $R_{pPb}$ for prompt $\psi(2S)$ production remains below unity independent of rapidity, while in the highest $p_T$ bin ($10 < p_T < 30$ GeV/c) it is consistent with unity (although systematically smaller). The $R_{pPb}$ values of prompt J/$\psi$ lie systematically above those of prompt $\psi(2S)$ mesons, indicating different nuclear effects in the production of the two states. The effects of nuclear parton distribution functions or coherent energy loss are expected to affect the $R_{pPb}$ of prompt J/$\psi$ and $\psi(2S)$ by a similar amount, thus the results hint the presence of final-state interactions with the medium produced in pPb collisions.